\documentclass[letterpaper]{article}

\usepackage{thumbpdf,lmodern}
\usepackage{amsmath,amssymb}
\usepackage{natbib}
\usepackage{hyperref}
\usepackage{rotating}
\usepackage[para]{threeparttable}
\usepackage{booktabs}
\usepackage{multirow}
\usepackage{makecell}
\usepackage{authblk}
\let\code=\texttt
\let\proglang=\textsf
\newcommand{\pkg}[1]{{\fontseries{b}\selectfont #1}}

\providecommand{\keywords}[1]
{
  \small	
  \textbf{\textit{Keywords---}} #1
}

\usepackage{fancyhdr}
\author[1]{Seyoon Ko}
\author[2]{Hua Zhou}
\author[3]{Jin Zhou}
\author[1]{Joong-Ho Won}
\affil[1]{Seoul National University}
\affil[2]{University of California, Los Angeles}
\affil[3]{University of Arizona}

\title{\pkg{DistStat.jl}: Towards Unified Programming for High-Performance Statistical Computing Environments in \proglang{Julia}}
\begin{document}
\maketitle

\begin{abstract}
  The demand for high-performance computing (HPC) is  ever-increasing for everyday statistical computing purposes.
  The downside is that we need to write specialized code for each HPC environment. 
  CPU-level parallelization needs to be explicitly coded for effective use of multiple nodes in cluster  supercomputing environments. 
  Acceleration via graphics processing units (GPUs) requires to write kernel code. 
  The \proglang{Julia} software package \pkg{DistStat.jl} implements a data structure for distributed arrays that work on both multi-node CPU clusters and multi-GPU environments transparently. 
  This package paves a way to developing high-performance statistical software in various HPC environments simultaneously.  
  As a demonstration of the transparency and scalability of the package, we provide applications to large-scale nonnegative matrix factorization, multidimensional scaling, and $\ell_1$-regularized Cox proportional hazards model on an 8-GPU workstation and a 720-CPU-core virtual cluster in Amazon Web Services (AWS) cloud. 
  As a case in point, we analyze the on-set of type-2 diabetes from the UK Biobank with 400,000 subjects and 500,000 single nucleotide polymorphisms using the $\ell_1$-regularized Cox proportional hazards model.
  Fitting a half-million-variate regression model took less than 50 minutes on AWS.
\end{abstract}


\section{Introduction} \label{sec:intro}
There is no doubt that statistical practice needs more and more computing power.
In recent years, increase in computing power is usually achieved by using more computing cores or interconnecting more machines. 
Modern supercomputers are dominantly cluster computers that utilize multiple cores of central processing units (CPUs) over multiple machines connected via fast communication devices. 
Also, co-processors such as graphics processing units (GPUs) are now widely used for accelerating many computing tasks involving linear algebra and convolution operations. 
In addition, as the cloud computing technology matures, users can now access virtual clusters through cloud service providers such as Amazon Web Services, Google Compute Platform, and Microsoft Azure without need of purchasing or maintaining machines physically. 
With the demand for analysis of terabyte- or petabyte-scale data in diverse disciplines, the crucial factor for the success of large-scale data analysis is how well one can utilize high-performance computing (HPC) environments.

However, the statistical community appears yet to fully embrace the power of HPC. 
This is partly because of the difficulty of programming in multiple nodes and on GPUs in \proglang{R}, the most popular programming language among statisticians. 
Furthermore, statisticians often face the burden of writing separate codes for different HPC environments. 
While there are a number of software packages in high-level languages, including \proglang{R}, that simplify GPU programming and multi-node programming separately (reviewed in Section \ref{sec:related}), those that enable simplification for both multi-GPU programming and multi-node programming with the unified code base are rare. 
In particular, there are few packages for multi-device distributed linear algebra with non-CPU environments. This leads to the necessity of a tool that merges programming and implement linear algebra operations for disparate HPC environments. High-performance implementation of easily parallelizable algorithms in statistical computing, for example, MM algorithms \citep{lange2000optimization,hunter2004tutorial,lange2016mm}, proximal gradient methods \citep{Beck:ISTA}, and proximal distance algorithms \citep{keys2019proximal} will benefit from such a tool.

In this paper, we introduce \pkg{DistStat.jl}, which implements a distributed array data structure on both distributed CPU and GPU environments and also provides an easy-to-use interface to the structure in the programming language \proglang{Julia} \citep{bezanson2017julia}. A user can switch between the underlying array implementations for CPU cores or GPUs only with minor configuration changes. Furthermore, \pkg{DistStat.jl} can generalize to any environment on which the message passing interface (\pkg{MPI}) is supported. This package leverages on the built-in support for multiple dispatch and vectorization semantics of \proglang{Julia}, resulting in easy-to-use syntax for elementwise operations and distributed linear algebra.  We also demonstrate the merits of \pkg{DistStat.jl}, in applications to highly-parallelizable statistical optimization problems. Concrete examples include nonnegative matrix factorization, multidimensional scaling, and $\ell_1$-regularized Cox regression. These examples were considered in \citet{ko2020high} with their \pkg{PyTorch} \citep{paszke2019pytorch} implementation, \pkg{dist\_stat}, of the algorithms. 
We selected these in order to 
to highlight the merits of \proglang{Julia} in developing high-performance statistical computing software. The improved performance brings sharp contrast in the genome-wide survival analysis of the UK Biobank data with the full 500,000 genomic loci and 400,000 subjects in a cloud of twenty 144 GB-memory, 36-physical-core instances. This doubles the number of subjects analyzed in \citet{ko2020high}.

The remainder of this paper is organized as follows. We review related software in Section \ref{sec:related}. Then, basic usage and array semantics of \proglang{Julia} are covered in Section \ref{sec:julia}. In Section \ref{sec:interface}, we review software interface of the package \pkg{DistStat.jl}. Section \ref{sec:applications} shows how \pkg{DistStat.jl} can be used with easily-parallelizable statistical optimization algorithms. We also compare the performance of \pkg{DistStat.jl} in \proglang{Julia} to that of \pkg{dist\_stat} in \proglang{Python}. We analyze the real-world UK Biobank data with the Cox proportional hazards model in Section \ref{sec:ukbiobank}. 
We conclude the paper with summary and discussion in Section \ref{sec:summary}.
The code is available at \url{https://github.com/kose-y/DistStat.jl}, released under the MIT License. The package can be installed using the instructions provided in Appendix \ref{sec:install}.

\section{Related Software}\label{sec:related}
\subsection{Message-passing interface and distributed array interfaces}
The de facto standard for inter-node communication in distributed computing environments is the message passing interface (\pkg{MPI}). The latter defines several ways to communicating between two processes (point-to-point communication) or a group of processes (collective communication). Although \pkg{MPI} is originally defined in \proglang{C} and \proglang{Fortran},  many other high-level languages 
offer wrappers to this interface:
\pkg{Rmpi} \citep{rmpi} for \proglang{R}, \pkg{mpi4py} \citep{mpi4py} for \proglang{Python}, and \pkg{MPI.jl} \citep{mpijl} for \proglang{Julia}, etc.

Leveraging on distributed computing environments,
there have been several attempts to incorporate array and linear algebra operations through the basic syntax of the base programming language. \proglang{MATLAB} has a distributed array implementation that uses \pkg{MPI} as a backend in the \pkg{Parallel Computing Toolbox}. In \proglang{Julia}, \pkg{MPIArrays.jl} \citep{mpiarraysjl} defines a matrix-vector multiplication routine that uses \pkg{MPI} as its backend. \pkg{DistributedArrays.jl} \citep{distributedarraysjl} is a more general attempt to create a distributed array, allowing various communication modes, including Transmission Control Protocol/Internet Protocol (TCP/IP) and Secure Shell (SSH), and MPI. In \proglang{R}, a package called \pkg{ddR} \citep{ma2016dmapply} supports distributed array operations.

\subsection{Unified array interfaces for CPU and GPU}
For GPU programming, \pkg{CUDA} \proglang{C} for Nvidia GPUs and \pkg{OpenCL} for generic GPUs are by far the most widely used. \proglang{R} package \pkg{gputools} \citep{buckner2010gputools} is one of the earliest efforts to incorporate GPU in \proglang{R}. \pkg{PyCUDA} and \pkg{PyOpenCL} \citep{klockner2012pycuda} for \proglang{Python} and \pkg{CUDA.jl} (previously \pkg{CUDAnative.jl} and \pkg{CUDAdrv.jl}) \citep{besard2018juliagpu} for \proglang{Julia} allow users to access low-level features of the respective interfaces. 

An interface to array and linear algebra operations that works transparently on both CPU and GPU is highly desirable, as it reduces the programming burden tremendously.
In \proglang{MATLAB}, the \pkg{Parallel Computing Toolbox} includes the data structure \code{gpuArray}. Simply wrapping an ordinary array with the function \code{gpuArray()} allows the users to use predefined functions to exploit the single instruction, multiple data (SIMD) 
parallelism
of Nvidia GPUs.
In Python, the recent deep learning \citep{lecun2015deep} boom accelerated development of easy-to-use array interfaces and linear algebra operations along with automatic differentiation for both CPU and GPU. The most popular among them are \pkg{TensorFlow} \citep{abadi2016tensorflow} and \pkg{PyTorch} \citep{paszke2019pytorch}. 
The latter development is worth noting since most statistical computing tasks, not alone neural networks, can benefit from it.
For example,
the \pkg{Distributions} subpackage of \pkg{TensorFlow} \citep{dillon2017tensorflow} 
provides a convenient programming environment for Bayesian computing on GPUs,
such as stochastic gradient Monte Carlo Markov chain \citep{baker2018sgmcmc}. 
In \proglang{Julia}, \pkg{CUDA.jl} (previously \pkg{CuArrays.jl})\citep{besard2019prototyping,cudajl} defines many array operations and simple linear algebra routines using the same syntax as the base CPU arrays. 

\subsection{Previous work}
To lessen the burden of programming for different HPC environments, it is desirable to unify 
these
approaches and have a common syntax for 
within-node CPU/GPU computation and inter-node/inter-GPU communication. 
In this regard, the authors have previously developed a package \pkg{dist\_stat} \citep{ko2020high} that implements distributed array and linear algebra operations using \pkg{PyTorch}.
Through the examples of nonnegative matrix factorization, multidimensional scaling, and $\ell_1$-regularized Cox regression, they have demonstrated the scalability of their approach on both multiple GPUs and multiple CPU nodes on a cloud.
This package provides a transparent and easy-to-use interface for 
distributed arrays
on many computing environments, e.g.,
a single CPU-only node, a single node with a single GPU, a single node with multiple GPUs, multiple nodes with only CPUs, and multiple nodes with multiple GPUs. 
However, there are a couple of limitations in \pkg{dist\_stat}, mainly coming from those of \pkg{PyTorch}. First, all the processes must have data of an equal size, 
hence the number of processes to be created must divide the total size of the data. This limite the tunability of the computation jobs.
Second, it is very difficult to write effective device-specific code for further acceleration without writing code in \proglang{C} or other packages that support just-in-time (JIT) compilation.
Hence the authors  decided to migrate to \proglang{Julia} in order to avoid these issues and leverage higher flexibility.



\section{Julia Basics} \label{sec:julia}
\proglang{Julia} is a high-level programming language that has a flavor of interpreter languages such as \proglang{R} and \proglang{Python}, but compiles for efficient execution via LLVM \citep{lattner2004llvm}. Its syntax is similar to those of \proglang{MATLAB} and \proglang{R}, leading to easy-to-read code that can run on various hardware with only minor changes, including CPUs and GPUs. In this section, we review the basic syntax of \proglang{Julia}. 
Our description is based on \proglang{Julia}  version 1.4. For more details, see the official documentation \citep{juliadoc}. 

\subsection{Methods and multiple dispatch}
In \proglang{Julia}, a function is ``an object that maps a tuple of argument values to a return value''. A function can have many different specific implementations, distinguished by the types of input arguments. 
Each specific implementation is called a \textbf{method},
and a method of a function is selected by the mechanism called \textbf{multiple dispatch}.
Many core functions in \proglang{Julia} possess several methods attached to each of them. 
A user can also define additional methods to existing functions.
For example, a method for function \code{f} can be defined as follows:
\begin{verbatim}
julia> f(x, y) = "foo"
f (generic function with 1 method)
\end{verbatim}
Each argument can be constrained to certain type, for example:
\begin{verbatim}
julia> f(x::Float64, y::Float64) = x * y
f (generic function with 2 methods)

julia> f(x::String, y::String) = x * y
f (generic function with 3 methods)
\end{verbatim}
\code{Float64} is the data type for a double-precision (64-bit) floating point number. An asterisk (\code{*}) between two \code{String} objects means concatenating them.
At runtime, the most specific method is used for the given combination of input arguments.
\begin{verbatim}
julia> f("Candy", 3.0)
"foo"

julia> f("test", "me")
"testme"

julia> f(2.0, 3.0)
6.0
\end{verbatim}
Methods and types may have parameters, enclosed by a pair of curly braces (\code{\{\}}). A parametric method is defined as follows:
\begin{verbatim}
julia> g(x::T, y::T) where {T <: Real} = x * y
g (generic function with 1 method)
\end{verbatim}
The function \code{g()} performs multiplication of the two arguments if the two arguments are the same subtype of \code{Real} (a type for real numbers, for example, \code{Float64}, \code{Int32} (32-bit integer), etc.) and both are of the same type. 
\begin{verbatim}
julia> g(2.0, 3.0)
6.0

julia> g(2, 3)
6

julia> g(2.0, 3)
ERROR: MethodError: no method matching g(::Float64, ::Int64)
Closest candidates are:
  g(::T<:Real, ::T<:Real) where T<:Real at REPL[17]:1
Stacktrace:
 [1] top-level scope at REPL[28]:1
\end{verbatim}
The third command throws an error, because the two arguments have different types. Such an error can be avoided by defining a more general method:
\begin{verbatim}
julia> g(x::Real, y::Real) = x * y
g (generic function with 2 methods)
\end{verbatim}
Here, the exact type of \code{x} and \code{y} may be different. An example of parametric types,  \code{AbstractArray}, is discussed in Section \ref{sec:array}. 

\subsection{Multidimensional arrays}\label{sec:array}
An array in \proglang{Julia} is defined as ``a collection of objects stored in a multi-dimensional grid''. Each object should be of a specific type for optimized performance, such as \code{Float64}, \code{Int32}, or \code{String}. 

The top-level abstract type for a multidimensional array is \code{AbstractArray\{T,N\}}, where parameter \code{T} is the type of element (such as \code{Float64}, \code{Int32}), and \code{N} is the number of dimensions. \code{AbstractVector\{T\}} and \code{AbstractMatrix\{T\}} are aliases for \code{AbstractArray\{T, 1\}} and \code{AbstractArray\{T, 2\}}, respectively. Operations for \linebreak \code{AbstractArray}s are provided as fallback methods which would generally work correctly in many cases, but are often slow. 

The type \code{DenseArray} is a subtype of \code{AbstractArray} representing an array stored in contiguous CPU memory. 
The most frequently used instance of  \code{DenseArray} is \code{Array}, a type for basic CPU array with a grid structure. \code{Vector\{T\}} and \code{Matrix\{T\}} are aliases for \code{Array\{T, 1\}} and \code{Array\{T, 2\}}, respectively. 
Another subtype of \code{DenseArray} is \code{CuArray}, defined in \pkg{CUDA.jl}, a contiguous array data type on a \pkg{CUDA} GPU. Many of array operations for \code{CuArray} are provided using the same syntax as \code{Array}s.

A \code{Matrix} (or an instance of \code{Array\{T, 2\}}) is easily created using a \proglang{MATLAB}-like syntax such as:
\begin{verbatim}
julia> A = [1 2; 3 4]
2×2 Array{Int64,2}:
 1  2
 3  4
\end{verbatim}
An \code{Array} can be allocated with undefined values using:
\begin{verbatim}
julia> B = Array{Float64}(undef, 2, 3)
2×3 Array{Float64,2}:
 6.90922e-310  6.90922e-310  6.90922e-310
 6.90922e-310  6.90922e-310  6.90921e-310
\end{verbatim}
There are predefined basic functions for array operations, such as \code{size(A)} that returns a tuple of dimensions of \code{A}, \code{eltype(A)} that returns the type of elements in \code{A}, and \code{ndims(A)}, that shows the number of dimensions of \code{A}. 

\subsection{Matrix multiplication} \label{sec:juliamatmul}
Linear algebra operations in \proglang{Julia} are defined in the basic package \code{LinearAlgebra}. The functions in \code{LinearAlgebra} can be loaded to the workspace with the keyword \code{using}:
\begin{verbatim}
julia> using LinearAlgebra
\end{verbatim}
Matrix multiplication in \proglang{Julia} is defined in the function \code{LinearAlgebra.mul!(C, A, B)}\footnote{It is a convention in \proglang{Julia} to end the name of a function that changes the value of its arguments with an exclamation mark (\code{!}).}. This function computes the post-multiplication of matrix \code{B} to matrix \code{A}, and stores the result in matrix \code{C}. The most general definition of \code{LinearAlgebra.mul!()} is:
\begin{verbatim}
LinearAlgebra.mul!(C::AbstractMatrix, A::AbstractVecOrMat, 
                    B::AbstractVecOrMat)
\end{verbatim}
which implements a naive algorithm for matrix multiplication. For a \code{Matrix} stored in the CPU memory, the call to \code{LinearAlgebra.mul!()} with arguments 
\begin{verbatim}
LinearAlgebra.mul!(C::Matrix, A::Matrix, B::Matrix)
\end{verbatim}
invokes the \code{gemm} (general matrix multiplication) routine of the \pkg{BLAS}, or the basic linear algebra subprograms \citep{blackford2002updated}, e.g., 
\begin{verbatim}
julia> A= [1. 2.; 3. 4.]
2×2 Array{Float64,2}:
 1.0  2.0
 3.0  4.0

julia> B = [3. 4.; 5. 6.]
2×2 Array{Float64,2}:
 3.0  4.0
 5.0  6.0

julia> C = Array{Float64, 2}(undef, 2, 2)
2×2 Array{Float64,2}:
 6.90922e-310  6.90921e-310
 6.90922e-310  6.90922e-310

julia> mul!(C, A, B)
2×2 Array{Float64,2}:
 13.0  16.0
 29.0  36.0

julia> C
2×2 Array{Float64,2}:
 13.0  16.0
 29.0  36.0
\end{verbatim}

On the other hand, for matrices on GPU, a call to the same function \linebreak \code{LinearAlgebra.mul!()} with arguments
\begin{verbatim}
LinearAlgebra.mul!(C::CuMatrix, A::CuMatrix, B::CuMatrix)
\end{verbatim}
results in operations using \pkg{cuBLAS} \citep{cublas}, 
a \pkg{CUDA} version of \pkg{BLAS}:
\begin{verbatim}
julia> using CUDA

julia> A_d = cu(A)
2×2 CuArray{Float32,2,Nothing}:
 1.0  2.0
 3.0  4.0

julia> B_d = cu(B)
2×2 CuArray{Float32,2,Nothing}:
 3.0  4.0
 5.0  6.0

julia> C_d = cu(C)
2×2 CuArray{Float32,2,Nothing}:
 0.0  0.0
 0.0  0.0

julia> mul!(C_d, A_d, B_d)
2×2 CuArray{Float32,2,Nothing}:
 13.0  16.0
 29.0  36.0
\end{verbatim}
The function \code{cu()} transforms an \code{Array\{T, N\}} into a \code{CuArray\{Float32, N\}}, transferring data from CPU to GPU.

\subsection{Dot syntax for vectorization}
\proglang{Julia} has a special ``dot'' syntax for vectorization. The dot syntax is invoked by prepending a dot to an operator (e.g., \code{.+}) or appending a dot to a function name (e.g., \code{soft\_threshold.()}). Unlike many other programming languages (e.g., \proglang{R}), vectorization in \proglang{Julia} can be applied to any function without a need to deliberately tailor the corresponding method. \proglang{Julia}'s JIT compiler automatically matches singleton dimensions of array arguments to the dimensions of other array arguments. For example,
\begin{verbatim}
julia> a = [1, 2]
2-element Array{Int64,1}:
 1
 2

julia> b = [3 4; 5 6]
2×2 Array{Int64,2}:
 3  4
 5  6

julia> a .+ b
2×2 Array{Int64,2}:
 4  5
 7  8
\end{verbatim}
Note that \code{a} is a column vector and \code{b} is a matrix. 

The dot syntax can be extended by defining the method \code{broadcast()} for each array interface, allowing its generalization to any underlying hardware architecture. In addition, multiple dots on the same line of code fuse into a single call to \code{broadcast()}, translated into a single vectorized loop (for CPU) or a single generated kernel (for GPU) for that line.

While broadcasting is one of the simplest way to specifying elementwise operations, it is often not the fastest option. 
Broadcasting often allocates excessive memory, thus well-optimized compiled loops without memory allocation may be faster in many cases.

\section{Software interface} \label{sec:interface}
In this section, we review the features of \pkg{DistStat.jl}. \pkg{DistStat.jl} implements a distributed \pkg{MPI}-based array data structure \code{MPIArray} 
as a subtype of \code{AbstractArray},
using \pkg{MPI.jl} as a backend. 
It has been tested for basic \code{Array}s and \code{CuArray}s. The standard vectorized ``dot'' operations can be used for convenient element-by-element operations on \code{MPIArray}s as well as broadcasting operations. Furthermore, key distributed matrix operations, such as the matrix-matrix multiplication, for \code{MPIMatrix}, or two-dimensional \code{MPIArray}s, are also implemented. Reduction and accumulation operations are supported for \code{MPIArray}s of any dimensions.   The package can be loaded by:
\begin{verbatim}
using DistStat
\end{verbatim}
If GPUs are available, one that is to be used is automatically selected in a round-robin fashion upon loading the package. The rank, or the ``ID'' of a process, and the size, or the total number of the processes, can be accessed by:
\begin{verbatim}
DistStat.Rank()
DistStat.Size()
\end{verbatim}
Ranks are indexed 0-based, following the \pkg{MPI} standard.
(Note \proglang{Julia}'s array indexes are 1-based.)

\subsection{Data Structure for Distributed MPI Array}
In \pkg{DistStat.jl}, a distributed array data type \code{MPIArray\{T,N,AT\}} is defined. Here, parameter \code{T} is the type of the elements of the array, e.g., \code{Float64} or \code{Float32}. Parameter \code{N} is the dimension of the array, \code{1} for vector and \code{2} for matrix, etc. Parameter \code{AT} (for ``array type'') is the implementation of \code{AbstractArray} used for base operations: \code{Array} for the basic CPU array, and  \code{CuArray} for the arrays on Nvidia GPUs (which require \pkg{CUDA.jl}). If there are multiple CUDA devices, a device is assigned to a process automatically by the rank of the process modulo the size. This assignment scheme extends to the setting in which there are multiple GPU devices in multiple CPU nodes. The type \code{MPIArray\{T,N,AT\}} is a subtype of \code{AbstractArray\{T, N\}}. 
In \code{MPIArray\{T,N,AT\}}, each rank holds a contiguous block of the full data in \code{AT\{T,N\}} split and distributed along the \code{N}-th or the last dimension of an \code{MPIArray}.

In the special case of two-dimensional arrays, aliased  \code{MPIMatrix\{T,AT\}} and created using the function \code{transpose()}, the data is column-major ordered and column-split. The transpose of this matrix has type of
\code{Transpose\{T,MPIMatrix\{T,AT\}\}},
which is a lazy wrapper representing row-major ordered and row-split matrix.
There also is an alias for one-dimensional array \code{MPIArray\{T,1,AT\}}, which is \code{MPIVector\{T,AT\}}.

\subsubsection{Creation}
The syntax \code{MPIArray\{T,N,AT\}(undef, m, ...)} creates an uninitialized \code{MPIArray}. For example,
\begin{verbatim}
a = MPIArray{Float64, 2, Array}(undef, 3, 4)
\end{verbatim}
creates an uninitialized 3$\times$4 distributed array based on local \code{Array}s of double precision floating-point numbers. The size of this array, the type of each element, and number of dimensions can be accessed using the usual functions in \proglang{Julia}: \code{size(a)}, \code{eltype(a)}, and \code{ndims(a)}. 
Local data held by each process can be accessed by appending \code{.localarray} to the name  of the array, e.g.,
\begin{verbatim}
a.localarray
\end{verbatim}
Matrices are split and distributed as evenly as possible. For example, if the number of processes is 4 and the \code{size(a) == (3, 7)}, processes of ranks 0 through 2 hold the local data of size (3, 2) and the rank-3 process holds the local data of size (3, 1).

An \code{MPIArray} can also be created by distributing an array residing in a single process. For example, in the following code:
\begin{verbatim}
if DistStat.Rank() == 0
    dat = [1, 2, 3, 4]
else
    dat = Array{Int64}(undef, 0)
end
d = distribute(dat)
\end{verbatim}
the data is defined in the rank-0 process, and each other process has an empty instance of \code{Array\{Int64\}}. Using the function \code{distribute}, the \code{MPIArray\{Int64, 1, Array\}} of the data \code{[1, 2, 3, 4]}, equally distributed over four processes, is created.

\subsubsection{Filling an array}
An \code{MPIArray} \code{a} can be filled with a number \code{x} using the usual syntax of the function \code{fill!(a, x)}. For example, \code{a} can be filled with zero:
\begin{verbatim}
fill!(a, 0)
\end{verbatim}

\subsubsection{Random number generation}
An array can also be filled with random values, extending \code{Random.rand!()} for the standard uniform distribution and \code{Random.randn!()} for the standard normal distribution. The following code fills \code{a} with uniform(0, 1) random numbers:
\begin{verbatim}
using Random
rand!(a)
\end{verbatim}
In cases such as unit testing, generating identical data for any configuration is important. For this purpose, the following interface is defined:
\begin{verbatim}
function rand!(a::MPIArray{T,N,A}; seed=nothing, common_init=false, 
    root=0) where {T,N,A}
\end{verbatim}
If the keyword argument \code{common\_init} is set \code{true}, the data are generated from the process with rank \code{root}. If \code{common\_init == false}, each process independently fills its local array with random numbers. A \code{seed} can also be passed. If \code{common\_init == false} and \code{seed == k}, the seed for each process is set to \code{k} plus the rank.

\subsection{``Dot'' syntax and vectorization}\label{sec:dot}
The ``dot'' broadcasting feature of \pkg{DistStat.jl} follows the standard \proglang{Julia} syntax. This syntax provides a convenient way to operate on both multi-node clusters and multi-GPU workstations with the same code. For example, the soft-thresholding operator, which appears commonly in sparse regression, can be defined in the element level: 
\begin{verbatim}
function soft_threshold(x::T, lambda::T)::T where T <: AbstractFloat
    x > lambda && return (x - lambda)
    x < -lambda && return (x + lambda)
    return zero(T)
end
\end{verbatim}
This function can be applied to each element of an \code{MPIArray} using the dot broadcasting, as follows. When the dot operation is used for an \code{MPIArray\{T,N,AT\}}, it is naturally passed to inner array implementation \code{AT}. Consider the following arrays filled with random numbers from the standard normal distribution:
\begin{verbatim}
a = MPIArray{Float64, 2, Array}(undef, 2, 4) |> randn!
b = MPIArray{Float64, 2, Array}(undef, 2, 4) |> randn!
\end{verbatim}
The function \code{soft\_threshold()} is applied elementwisely as the following:
\begin{verbatim}
a .= soft_threshold.(a .+ 2 .* b, 0.5)
\end{verbatim}
The three dot operations, \code{.=}, \code{.+}, and \code{.*}, are fused into a single loop (in CPU) or a single kernel (in GPU) internally. 

A singleton non-last dimension is treated as if the array is repeated along that dimension, just like \code{Array} operations. For example,
\begin{verbatim}
c = MPIArray{Float64, 2, Array}(undef, 1, 4) |> rand!
a .= soft_threshold.(a .+ 2 .* c, 0.5)
\end{verbatim}
works as if \code{c} were a $2 \times 4$ array, with its content repeated twice.
The terminal dimension is a little bit subtle, as an \code{MPIArray\{T,N,AT\}} is distributed along that dimension. 
Dot broadcasting along this direction works if the broadcast array has the type \code{AT} and holds the same data across the processes. For example, 
\begin{verbatim}
d = Array{Float64}(undef, 2, 1); fill!(d, -0.1)
a .= soft_threshold.(a .+ 2 .* d, 0.5)
\end{verbatim}
As with any dot operations in \proglang{Julia}, the dot operations for \pkg{DistStat.jl} are convenient but usually not the fastest option. Its implementations can be further optimized by specializing in specific array types. An example of this is given in Section \ref{sec:cox}.

\subsection{Reduction operations and accumulation operations}
Reduction operations, such as \code{sum()}, \code{prod()}, \code{maximum()}, \code{minimum()}, and accumulation operations, such as \code{cumsum()}, \code{cumsum!()}, \code{cumprod()}, \code{cumprod!()}, are implemented just like their base counterparts. 
Example usages of \code{sum()} and \code{sum!()} are:
\begin{verbatim}
sum(a)
sum(abs2, a) 
sum(a, dims=1)
sum(a, dims=2)
sum(a, dims=(1,2)) 
sum!(c, a) 
sum!(d, a) 
\end{verbatim}
The first line computes the elementwise sum of \code{a}. The second line computes the sum of squared absolute values (\code{abs2()} is the function that computes the squared absolute values). The third and fourth lines compute the column sums and row sums, respectively. Similar to the dot operations, the third line reduces along the distributed dimensions, and returns a broadcast local \code{Array}. The fifth line returns the sum of all elements, but the data type is a $1 \times 1$ \code{MPIArray}. The syntax \code{sum!(p, q)} selects which dimension to reduce based on the shape of \code{p}, the first argument. The sixth line computes the columnwise sum and saves it to \code{c}, because \code{c} is a $1 \times 4$ \code{MPIArray}. The seventh line computes rowwise sum, because \code{d} is a $2 \times 1$ local \code{Array}.  

Given below are examples for \code{cumsum()} and \code{cumsum!()}:
\begin{verbatim}
cumsum(a; dims=1) 
cumsum(a; dims=2) 
cumsum!(b, a; dims=1)
cumsum!(b, a; dims=2)
\end{verbatim}
The first line computes the columnwise cumulative sum, and the second line computes the rowwise cumulative sum. So do the third and fourth lines, but save the results in \code{b}, which has the same size as \code{a}. 
\subsection{Distributed linear algebra}\label{sec:distlinalg}
\subsubsection{Dot product}
The method \code{LinearAlgebra.dot()} for \code{MPIArray}s is defined just like the base \linebreak \code{LinearAlgebra.dot()}, which sums all the elements after an elementwise multiplication of the two argument arrays: 
\begin{verbatim}
using LinearAlgebra
dot(a, b)
\end{verbatim}
\subsubsection{Operations on the diagonal}
The ``getter'' method for the diagonal, \code{diag!(d, a)}, and the ``setter'' method for the diagonal, \code{fill\_diag!()}, are also available.
The former obtains the main diagonal of the \code{MPIMatrix} \code{a} and is stored in \code{d}. If \code{d} is an \code{MPIMatrix} with a single row, the result is obtained in a distributed form. On the other hand, if \code{d} is a local \code{AbstractArray}, all elements of the main diagonal is copied to all processes as a broadcast \code{AbstractArray}:
\begin{verbatim}
M = MPIMatrix{Float64, Array}(undef, 4, 4) |> rand!
v_dist = MPIMatrix{Float64, Array}(undef, 1, 4)
v = Array{Float64}(undef, 4)
diag!(v_dist, M)
diag!(v, M)
\end{verbatim}

\subsubsection{Matrix multiplication}
Function \code{LinearAlgebra.mul!(C, A, B)} left-multiplies matrix \code{A} ($p \times q$) to another matrix \code{B} ($q \times r$) to store the product to matrix \code{C} ($p \times r$).
The present implementation of \code{mul!()} for \code{MPIMatrix}es improves the six distributed matrix multiplication scenarios for multiplying tall-and-skinny and/or wide-and-short matrices implemented  using Python and \pkg{MPI} in \citet{ko2020high}. 
Each input matrix can be either broadcast (represented by the type \code{AbstractMatrix}), column-split-and-distributed (\code{MPIMatrix}), or row-split-and-distributed (transposed  \code{MPIMatrix}), resulting in nine possible combinations.
Since \pkg{MPI} communication is not needed when both 
\code{A} and \code{B} are broadcast, there remain eight 
possibilities.
While the configuration of the output \code{C} is mostly determined by the input combination, there are cases in which multiple configurations are possible. 
Table \ref{tab:mul} collects the scenarios considered.
If \code{A} is an \code{MPIMatrix} (column-distributed) and \code{B} has the same configuration or is broadcast, then the
output \code{C} is a \code{MPIMatrix} to preserve the row distribution of \code{A} (Scenarios \textit{a} and \textit{e}).
However, if \code{B} is a transposed \code{MPIMatrix} (row-distributed), then
\code{C} can be any of \code{AbstractMatrix}, \code{MPIMatrix}, or the transposed \code{MPIMatrix}  (Scenarios \textit{b}, \textit{c}, \textit{d}).
Likewise,
if \code{A} is a transposed \code{MPIMatrix} 
and \code{B} has the same configuration or is broadcast, then the output \code{C} is a transposed \code{MPIMatrix} preserving the row distribution of \code{A} (Scenarios \textit{h} and \textit{i}),
while if 
\code{B} is a \code{MPIMatrix}, then \code{C} can be either an \code{MPIMatrix} or a transposed \code{MPIMatrix} (Scenarios \textit{f} and \textit{g}). 
When \code{A} is broadcast but \code{B} is distributed, the splitting of the inner dimension $q$ does not affect the that of the output (Scenarios \textit{j} and \textit{k}).
Thus, there are total 11 scenarios of matrix-matrix multiplication. 
%
%
Additional six scenarios are dedicated for
matrix-vector multiplication, where \code{A} is a \code{MPIMatrix} or its transpose, and \code{B} is either 
\code{MPIColVector\{T, AT\}} 
defined as
\code{Union\{MPIVector\{T,AT\}, Transpose\{T, MPIMatrix\{T,AT\}\}\}} 
to allow distribution of the rows,
or broadcast \code{AbstractVector}. 
Scenarios \textit{b} and \textit{c} for matrix-matrix multiplication merge into Scenario \textit{l}; Scenarios \textit{d} and \textit{e} translate into Scenarios \textit{m} and \textit{n}, respectively. The rest are similarly translated.

The internal implementation uses collective communication calls including \linebreak \code{MPI.Allreduce!()} and \code{MPI.Allgather!()} from \pkg{MPI.jl} for communication. 
For example, when \code{A} is an \code{MPIMatrix} ($p \times q$, column-split), \code{B} is a transposed \code{MPIMatrix} ($q \times r$, row-split), and \code{C} is a broadcast \code{AbstractMatrix} ($p \times r$), each process computes the local $p \times r$ multiplication first. Then, the results are reduced using \code{MPI.Allreduce!()} to compute the sum of local results.

In addition to the base syntax \code{mul!(C, A, B)}, 
an extra keyword argument for a temporary memory can be optionally provided, e.g, \code{mul!(C, A, B; tmp=Array(undef, 3, 4))}, 
in order to to save intermediate results.
This is useful for  avoiding repetitive allocations in iterative algorithms. The user should determine which shape of \code{C} minimizes communication and suits better for their application. 

\begin{sidewaystable}[]
\centering
\resizebox{\textwidth}{!}{%
\begin{tabular}{rllllll}
\hline
  & \code{A} ($p \times q$)                     & \code{B} ($q \times r$)                      & \code{C} ($p \times r$)                      & \code{tmp} (if defined) & dimension of \code{tmp} \\ \hline
a & \code{MPIMatrix} 
& \code{MPIMatrix} 
& \code{MPIMatrix} 
& \code{AbstractMatrix}   & $p \times q$                             \\
b & \code{MPIMatrix}                            & Transposed \code{MPIMatrix} 
& \code{MPIMatrix}                             & \code{AbstractMatrix}   & $p \times r$                             \\
c & \code{MPIMatrix}                            & Transposed \code{MPIMatrix}                  & Transposed \code{MPIMatrix} 
& \code{AbstractMatrix}   & $r \times p$                             \\
d & \code{MPIMatrix}                            & Transposed \code{MPIMatrix}                  & \code{AbstractMatrix}                        &                                          &                                          \\
e & \code{MPIMatrix}                            & \code{AbstractMatrix}                        & \code{AbstractMatrix}                        &                                          &                                          \\
f & Transpose \code{MPIMatrix} 
& \code{MPIMatrix}                             & \code{MPIMatrix}                             & \code{AbstractMatrix}   & $q \times p$                             \\
g & Transpose \code{MPIMatrix}                  & \code{MPIMatrix}                             & Transposed \code{MPIMatrix}                  & \code{AbstractMatrix}   & $q \times r$                             \\
h & Transpose \code{MPIMatrix}                  & Transposed \code{MPIMatrix}                  & Transposed \code{MPIMatrix}                  & \code{AbstractMatrix}   & $r \times q$                             \\
i & Transpose \code{MPIMatrix}                  & \code{AbstractMatrix}                        & Transposed \code{MPIMatrix}                  &                                          &                                          \\
j & \code{AbstractMatrix}                       & \code{MPIMatrix}                             & \code{MPIMatrix}                             &                                          &                                          \\
k & \code{AbstractMatrix}                       & Transposed \code{MPIMatrix}                  & \code{AbstractMatrix}                        &                                          &                                          \\
l & \code{MPIMatrix}                            & \code{MPIColVector}                          & \code{MPIColVector}                          & \code{AbstractVector}   & $p$                                      \\
m & \code{MPIMatrix}                            & \code{MPIColVector}                          & \code{AbstractVector}                        &                                          &                                          \\
n & \code{MPIMatrix}                            & \code{AbstractVector}                        & \code{AbstractVector}                        &                                          &                                          \\
o & Transposed \code{MPIMatrix}                 & \code{MPIColVector}                          & \code{MPIColVector}                          & \code{AbstractVector}   & $q$                                      \\
p & Transposed \code{MPIMatrix}                 & \code{AbstractVector}                        & \code{MPIColVector}                          &                                          &                                          \\
q & Transposed \code{MPIMatrix}                 & \code{AbstractVector}                        & \code{AbstractVector}                        &                                          &                                         \\ \hline
\end{tabular}
}
\caption{List of implementations for \code{LinearAlgebra.mul!(C, A, B)} with an optional keyword argument \code{tmp} for temporary storage.} 
\label{tab:mul}
\end{sidewaystable}

\subsubsection{Operator norms}
The method \code{opnorm()} either evaluates ($\ell_1$ and $\ell_\infty$) or approximates ($\ell_2$)  matrix operator norms, defined for a matrix $A \in \mathbb{R}^{m \times n}$ as $\|A\| = \sup\{\|Ax\|: x \in \mathbb{R}^n \text{ with } \|x\| = 1\}$ for each respective vector norm.
\begin{verbatim}
opnorm(a, 1)
opnorm(a, 2) 
opnorm(a, Inf)
\end{verbatim}
The $\ell_2$-norm is estimated via the power iteration \citep{golub2013matrix}, and can be further configured for convergence criterion and number of iterations. There also is an implementation based on the inequality $\|A\|_2 \le \|A\|_1 \|A\|_\infty$ (\code{method="quick"}), which overestimates the $\ell_2$-norm.
\begin{verbatim}  
opnorm(a, 2; method="power", tol=1e-6, maxiter=1000, seed=95376)
opnorm(a, 2; method="quick")
\end{verbatim}

\section{Applications}\label{sec:applications}
In this section, we consider several statistical applications implemented with \pkg{DistStat.jl}, namely the  nonnegative matrix factorization (NMF), multidimensional scaling (MDS), and $\ell_1$-regularized Cox proportional hazards regression. All of these applications require efficient iterative algorithms to scale up to high dimensions. For each mathematical description of the algorithms, we provide a code snippet for the  implementation. The snippets are simplified for ease of exposition, and the full implementation is given in the \code{examples/} directory in the code submitted. These code files should be run in such a way to incorporate \pkg{MPI}: by using \code{mpirun} or through the job scheduler. Also, we demonstrate the scalability of \pkg{DistStat.jl} over multiple GPUs on a local workstation and a virtual cluster on Amazon Web Services Elastic Compute Cloud (AWS EC2) with these examples and compare the timing with \pkg{dist\_stat}, our \pkg{PyTorch} implementation of distributed arrays \citep{ko2020high}. \proglang{Julia} 1.2.0 was used for \pkg{DistStat.jl}, and \pkg{PyTorch} 1.0 compiled from source installed on \proglang{Python} 3.6 was used for \pkg{dist\_stat}. To compare the execution time, we fixed the number of iterations. The system configurations are summarized in Table \ref{tab:setting}. These are deliberated set the same as \citet{ko2020high} for fair comparison. The virtual AWS EC2 cluster was managed by \pkg{CfnCluster} \citep{cfncluster}.

\begin{table}[]
\centering
\resizebox{\textwidth}{!}{
\begin{tabular}{lccc}
\hline
& \multicolumn{2}{c}{local node} & AWS c5.18xlarge \\
        \cmidrule(lr){2-3} \cmidrule(lr){4-4} 
             & CPU                   & GPU             & CPU\\ \hline
Model        & Intel Xeon E5-2680 v2 & Nvidia GTX 1080 & \begin{tabular}{@{}c@{}}Intel Xeon \\ Platinum 8124M\end{tabular} \\
\# of cores     & 10                    & 2560            & 18\\
Clock        & 2.8 GHz                 & 1.6 GHz           & 3.0GHz \\
\# of entities        & 2                     & 8               & \begin{tabular}{@{}c@{}}2 (per instance) \\ $\times$ 1-20 (instances)\end{tabular} \\
Total memory & 256 GB                 & 64 GB            & 144 GB $\times$ 4--20 \\
Total cores  & 20                    & 20,480 (CUDA)          & 36 $\times$ 4--20 \\ \hline
\end{tabular}
}
\caption{Hardware configuration of experiments. This table is equivalent to Table 2 of \citet{ko2020high}}\label{tab:setting}
\end{table}

All the experiments using AWS EC2 instances used double-precision floating-point numbers. All the GPU experiments were conducted with single-precision floating-point numbers unless otherwise noted. It is previously reported that the single-precision results are equivalent to the double-precision results up to six significant digits \citep{ko2020high}. 

In general, multi-GPU implementation results of \pkg{DistStat.jl} are largely comparable to that of \pkg{dist\_stat}. In large-scale AWS EC2 experiments, \pkg{DistStat.jl} achieved faster computation thanks to the increased flexibility of process configuration: when the communication is heavy, we can use a configuration with less jobs, each job using more threads; when communication is of little problem, we can use a configuration with more jobs, each job using a single thread. This is nearly impossible with \pkg{dist\_stat}: due to the limitation of \code{torch.distributed} subpackage of \pkg{PyTorch}, each process has to hold the same size of data. In addition, its \pkg{MPI} wrappers force copying of data before and after the data communication, while \pkg{MPI.jl} on the backend of \pkg{DistStat.jl} does not.

\subsection{Nonnegative matrix factorization}
Given a matrix $X \in \mathbb{R}^{m \times n}$ that consists of nonnegative values, NMF approximates $X$ by a product of two rank-$r$ ($r \ll \min\{m, n\}$) nonnegative matrices $V\in \mathbb{R}^{m \times r}$ and $W \in \mathbb{R}^{r \times n}$. This method has been applied in diverse disciplines, such as bioinformatics, recommender systems, astronomy, and image processing \citep{wang2013nonnegative}. In achieving this goal,  consider minimizing the cost function
\begin{align*}
    f(V, W) &= \| X - VW \|_\mathrm{F}^2,
\end{align*}
where $\| M \|_\mathrm{F}$ is the Frobenius norm of matrix $M$. A famous multiplicative algorithm to compute a local minimum of this cost function is due to \citet{lee1999learning,lee2001algorithms}:
\begin{align*}
V_{(n+1)} &= V_{(n)} \odot X W_{(n)}^\top \oslash V_{(n)} W_{(n)} W_{(n)}^\top \\
W_{(n+1)} &= W_{(n)} \odot V_{(n+1)}^\top X \oslash V_{(n+1)}^\top V_{(n+1)}^\top W_{(n)}, 
\end{align*}
where $\odot$ and $\oslash$ are elementwise multiplication and division, respectively. An alternative is  the alternating projected gradient (APG) algorithm \citep{lin2007projected}:
\begin{align*}
V_{(n+1)} &= \max\{0, V_{(n)} - \sigma_n (V_{(n)} W_{(n)} W_{(n)}^\top - X W_{(n)}^\top)\} \\
W_{(n+1)} &= \max\{0, W_{(n)} - \tau_n (V_{(n+1)}^\top V_{(n+1)} W_{(n)} - V_{(n+1)}^\top X)\},
\end{align*}
where the maximum is applied in an elementwise fashion. The $\sigma_n$ and $\tau_n$ are the step sizes, for which we can choose $\sigma_n = 1 / 2\|W_{(n)} W_{(n)}^\top\|_\mathrm{F}^2$ and $\tau_n = 1/2\|V_{(n+1)}^\top V_{(n+1)}\|_\mathrm{F}^2$. Such selection is known to promote the algorithm to converge in fewer iterations than the multiplicative algorithm \citep{ko2020high}. 

An implementation of the multiplicative algorithm of NMF in \pkg{DistStat.jl} is
\begin{verbatim}
using DistStat, Random, LinearAlgebra
m = 10000; n = 10000; r = 20
T = Float64; A = Array
X  = MPIMatrix{T, A}(undef, m, n); rand!(X)
Vt = MPIMatrix{T, A}(undef, r, m); rand!(Vt)
W  = MPIMatrix{T, A}(undef, r, n); rand!(W)
WXt = MPIMatrix{T, A}(undef, r, m)
WWt = A{T}(undef, r, r)
WWtVt = MPIMatrix{T, A}(undef, r, m)
VtX = MPIMatrix{T, A}(undef, r, n)
VtV = A{T}(undef, r, r)
VtVW = MPIMatrix{T, A}(undef, r, n)
eps = 1e-10
for i in 1:iter
    mul!(WXt, W, transpose(X)) 
    mul!(WWt, W, transpose(W))
    mul!(WWtVt, WWt, Vt)
    Vt .= Vt .* WXt ./ (WWtVt .+ eps)
    mul!(VtX, Vt, X)
    mul!(VtV, Vt, transpose(Vt))
    mul!(VtVW, VtV, W)
    W .= W .* VtX ./ (VtVW .+ eps)
end
\end{verbatim}
A main loop for the APG algorithm is given by:
\begin{verbatim}
for i in 1:iter
    mul!(WXt, W, transpose(X)) 
    mul!(WWt, W, transpose(W))
    mul!(WWtVt, WWt, Vt)
    sigma = 1.0 / (2 * (sum(WWt.^2)) + eps)
    Vt .= max.(0.0, Vt .- sigma .* (WWtVt .- WXt))
    mul!(VtX, Vt, X)
    mul!(VtV, Vt, transpose(Vt))
    mul!(VtVW, VtV, v.W)
    tau = 1.0 / (2 * (sum(VtV.^2)) + eps)
    W .= max.(0.0, W .- tau .* (VtVW .- VtX))
end
\end{verbatim}
This code runs on a single CPU node, and can be modified to run on a GPU workstation (multiple GPUs are allowed) by importing the package \pkg{CUDA.jl} and changing \code{A = Array} to \code{A = CuArray} on the third line. This selection is made through a command-line argument for the full implementation (in the \code{examples/} directory), where multi-node clusters are supported. 
Matrix $W$ (\code{W}) and the transpose of $V$ (\code{Vt}) are updated for each iteration. The intermediate results, \code{WXt} ($W X^\top$), \code{WWt} ($W W^\top$), \code{WWtVt} ($WW^\top V^\top$), \code{VtX} ($V^\top X$), \code{VtV} ($V^\top V$), and \code{VtVW} ($V^\top V W$) are preallocated. A very small number (\code{eps}) is added to the denominator for numerical stability. 

Thanks to the transparent implementation of distributed arrays and \code{mul!()} in Section \ref{sec:interface}, the main loop is not different from what one would write with native \proglang{Julia}. The same code with slight changes in the first three lines run on various HPC environments including CPU clusters and multi-GPU workstations in a distributed fashion. In our numerical experiments, further optimization for memory efficiency was conducted.

Table \ref{tab:nmfgpu} compares the performance of the two NMF algorithms on the multi-GPU setting in Table \ref{tab:setting} with $10,000 \times 10,000$ data for 10,000 iterations. It can be seen that the performance of the two algorithms is comparable, with APG being slightly slower with fixed number of iterations. This is because APG involves slightly more operations. With more than 4 GPUs, the communication burden outweighs the speed-up from using more GPU cores, and the algorithm slows down. The execution time between the \pkg{dist\_stat} and \pkg{DistStat.jl} implementations are also largely comparable, with \pkg{DistStat.jl} version being faster in $r=20$ cases. Experiments with 3, 6, or 7 GPUs were not possible with \pkg{dist\_stat}, because the size of the data was not divisible by 3, 6, and 7.

\begin{table}[]
\centering
\resizebox{\textwidth}{!}{
\begin{tabular}{lrrrrrrr} \hline
& \multirow{2}{*}{GPUs}              & \multicolumn{3}{c}{\pkg{dist\_stat}}      & \multicolumn{3}{c}{\pkg{DistStat.jl}}      \\
        \cmidrule(lr){3-5} \cmidrule(lr){6-8} 
&             & $r=20$    & $r=40$ & $r=60$ & $r=20$  & $r=40$ & $r=60$ \\ \hline
\multirow{8}{*}{Multiplicative  }  & 1              & 62      & 71   & 75   & 62    & 72   & 83   \\
& 2              & 43      & 55   & 63   & 42    & 60   & 72   \\
& 3              & --      & --   & --   & 38    & 57   & 71   \\
& 4              & 37      & 51   & 63   & 34    & 54   & 68   \\
& 5              & 39      & 54   & 66   & 38    & 56   & 75   \\
& 6              & --      & --   & --   & 36    & 56   & 80   \\
& 7              & --      & --   & --   & 37    & 58   & 81   \\
& 8              & 40      & 60   & 75   & 37    & 59   & 83   \\ \hline
\multirow{8}{*}{APG  }  & 1              & 68      & 76   & 82   & 61    & 80   & 85   \\
& 2              & 49      & 61   & 69   & 43    & 60   & 79   \\
& 3              & --      & --   & --   & 38    & 59   & 74   \\
& 4              & 44      & 58   & 70   & 36    & 54   & 72   \\
& 5              & 46      & 60   & 73   & 37    & 59   & 78   \\
& 6              & --      & --   & --   & 37    & 56   & 75   \\
& 7              & --      & --   & --   & 38    & 61   & 88   \\
& 8              & 47      & 68   & 83   & 39    & 59   & 82   \\ \hline
\end{tabular}
}
\caption{Runtime (in seconds) of NMF algorithms on $10,000 \times 10,000$ simulated data on GPUs. Experiments  could not be run with 3, 6, and 7 GPUs on \pkg{dist\_stat}, because 3, 6, and 7 did not divide the data size.}\label{tab:nmfgpu}
\end{table}

Table \ref{tab:nmfaws} compares the algorithms and implementations using $200,000 \times 200,000$ data on multiple AWS EC2 instances for 1000 iterations. We used two processes per instance to avoid the communication burden. Once again, elapsed time was largely similar between the two algorithms. APG was faster than the multiplicative algorithms in more cases compared to the GPU case, because the multiplicative algorithm on CPU often suffers from  slowdown due to creation of denormal numbers \citep{ko2020high}.  Between the two implementations, the \pkg{DistStat.jl} implementation was faster in 24 out of 30 cases. 

\begin{table}[]
\centering
\resizebox{\textwidth}{!}{
\begin{tabular}{lrrrrrrr} \hline
& \multirow{2}{*}{Instances}              & \multicolumn{3}{c}{\pkg{dist\_stat}}      & \multicolumn{3}{c}{\pkg{DistStat.jl}}      \\
        \cmidrule(lr){3-5} \cmidrule(lr){6-8} 
&            & $r=20$    & $r=40$ & $r=60$ & $r=20$  & $r=40$ & $r=60$ \\ \hline
\multirow{6}{*}{Multiplicative  }  & 4        & 1419           & 1748           & 2276           & 1392         & 1576         & 2057         \\
& 5        & 1076           & 1455           & 1698           & 1187         & 1383         & 1847         \\
& 8        & 859            & 966            & 1347           & 851          & 936          & 1430         \\
& 10       & 651            & 881            & 1115           & 708          & 856          & 1065         \\
& 16       & 549            & 700            & 959            & 553          & 694          & 907          \\
& 20       & 501            & 686            & 869            & 554          & 672          & 832          \\ \hline
\multirow{6}{*}{APG}  & 4        & 1333           & 1756           & 2082           & 1412         & 1711         & 2023         \\
& 5        & 1088           & 1467           & 1720           & 1215         & 1372         & 1775         \\
& 8        & 766            & 994            & 1396           & 849          & 916          & 1388         \\
& 10       & 677            & 870            & 1165           & 673          & 799          & 1014         \\
& 16       & 539            & 733            & 936            & 547          & 684          & 867          \\
& 20       & 506            & 730            & 919            & 538          & 727          & 836          \\ \hline
\end{tabular}
}
\caption{Runtime (in seconds) of NMF algorithms on $200,000 \times 200,000$ simulated data on multiple AWS EC2 instances}\label{tab:nmfaws}
\end{table}

\subsection{Multidimensional scaling}

MDS is a dimensionality reduction method that embeds a high-dimensional data  $X = [x_1, \dotsc, x_m]^\top \in \mathbb{R}^{m \times n}$ into $\boldsymbol{\theta} = [\theta_1, \dotsc, \theta_m]^\top \in \mathbb{R}^{m \times q}$ in a lower dimensional Euclidean space ($q \ll n$) while keeping the distance measures as close as possible. We use the cost function
\begin{align*}
f(\theta) &= \sum_{i=1}^m \sum_{j \ne i} w_{ij} (y_{ij} - \|\theta_i - \theta_j\|_2)^2 \\
          &= \sum_{i=1}^m \sum_{j \ne i} \left[ -2 w_{ij} y_{ij} \|\theta_i - \theta_j\|_2 + w_{ij} \|\theta_i - \theta_j\|_2^2 \right] + \mathrm{const.},
\end{align*}
where $y_{ij}$ is the distance measure between $x_i$ and $x_j$; and $w_{ij}$ are the weights. Using the majorization-minimization principle \citep{de1977convergence,deleeuw1977mds}, the update equation is given by 
\begin{align*}
\theta_{(n+1)ik} &= \left. \left(\sum_{j \ne i} w_{ij}\left[y_{ij} \frac{\theta_{(n)ik} - \theta_{(n)jk}}{\|\theta_{(n)i}- \theta_{(n)j}\|_2} + (\theta_{(n)ik} + \theta_{(n)jk})\right]\right) \right/ \left(2 \sum_{j \ne i} w_{ij} \right).
\end{align*}
for $i = 1, \dotsc, m$ and $k = 1, \dotsc, q$. The code below is a simple implementation of  MDS in \pkg{DistStat.jl}, which runs on the same environment as the example of the previous subsection.
\begin{verbatim}
using DistStat, Random, LinearAlgebra
m = 1000; n = 1000; r = 20
T = Float64; A = Array; iter=1000
X = MPIMatrix{T, A}(undef, m, n); rand!(X)
Y = MPIMatrix{T, A}(undef, n, n)
theta = MPIMatrix{T, A}(undef, r, n); rand!(theta)
theta = 2theta .- 1.0 # range (-1, 1)
theta_distances = MPIMatrix{T, A}(undef, n, n)
theta_WmZ = MPIMatrix{T, A}(undef, r, n)
d_dist = MPIMatrix{T, A}(undef, 1, n) # dist. row vector
d_local = A{T}(undef, n) # bcast. col vector
DistStat.euclidean_distance!(Y, X) # compute pairwise dist
W_sums = convert(T, n - 1)
for i in 1:iter
    mul!(theta_distances, transpose(theta), theta)
    diag!(d_dist, theta_distances)
    diag!(d_local, theta_distances)
    theta_distances .= -2theta_distances .+ d_dist .+ d_local
    fill_diag!(theta_distances, Inf)
    Z = Y ./ theta_distances
    Z_sums = sum(Z; dims=1) # Z sums, length n dist. row vector.
    WmZ = 1.0 .- Z
    fill_diag!(WmZ, zero(T)) # weights of diagonal elements are zero
    mul!(theta_WmZ, theta, WmZ)
    theta .= (theta .* (Z_sums .+ W_sums) .+ theta_WmZ) ./ 2W_sums
end
\end{verbatim}
This code can also run with local arrays if \code{diag!()} and \code{fill\_diag!()} are defined for them. The function \code{DistStat.euclidean\_distance!(Y, X)} computes the pairwise Euclidean distance matrix $Y$ between the input points in  $X$, which incorporates the chunking method \citep{li2010chunking}. 

Table \ref{tab:mdsgpu} compares the performance of \pkg{DistStat.jl} and \pkg{dist\_stat} on multiple GPUs with $10,000 \times 10,000$ data matrix $X$. Timing of \pkg{dist\_stat} was faster when the number of GPUs employed was small. This is because \pkg{dist\_stat} uses highly-optimized precompiled kernels from \pkg{PyTorch}, while \pkg{DistStat.jl} uses just-in-time compiled generated kernels. However, the gap between the two implementations vanishes as more GPUs are utilized.

\begin{table}[]
\centering
\begin{tabular}{rrrrrrr} \hline
\multirow{2}{*}{GPUs}               & \multicolumn{3}{c}{\pkg{dist\_stat}}      & \multicolumn{3}{c}{\pkg{DistStat.jl}}      \\
        \cmidrule(lr){2-4} \cmidrule(lr){5-7} 
           & $q=20$    & $q=40$ & $q=60$ & $q=20$  & $q=40$ & $q=60$ \\ \hline
1       & 292            & 301            & 307            & 402          & 415          & 423          \\
2       & 146            & 151            & 154            & 267          & 275          & 279          \\
3       & --             & --             & --             & 210          & 212          & 216          \\
4       & 81             & 84             & 88             & 89           & 93           & 97           \\
5       & 74             & 78             & 80             & 77           & 83           & 85           \\
6       & --             & --             & --             & 64           & 70           & 72           \\
7       & --             & --             & --             & 58           & 64           & 69           \\
8       & 52             & 58             & 64             & 53           & 60           & 65           \\ \hline
\end{tabular}
\caption{Runtime (in seconds) of MDS on $10,000 \times 10,000$ simulated data on multiple GPUs. Experiments  could not be run with 3, 6, and 7 GPUs on \pkg{dist\_stat}, because 3, 6, and 7 did not divide the data size.} \label{tab:mdsgpu}
\end{table}

For the AWS experiments, we used 36 processes per instance, because the step that mainly causes inter-instance communication is the matrix-vector multiplication, and its communication cost is much less than NMF. Note that this setting was not possible with the \pkg{dist\_stat} implementation. Table \ref{tab:mdsaws} shows the runtime of each experiment for 1000 iterations on $100,000 \times 1000$ dataset. It can be easily seen that \pkg{DistStat.jl} implementation is significantly faster.

\begin{table}[]
\centering
\begin{tabular}{rrrrrrr} \hline
\multirow{2}{*}{Instances}               & \multicolumn{3}{c}{\pkg{dist\_stat}}      & \multicolumn{3}{c}{\pkg{DistStat.jl}}      \\
        \cmidrule(lr){2-4} \cmidrule(lr){5-7} 
           & $q=20$    & $q=40$ & $q=60$ & $q=20$  & $q=40$ & $q=60$ \\ \hline
4        & 2875           & 3097           & 3089           & 2093         & 2007         & 2188         \\
5        & 2315           & 2378           & 2526           & 1625         & 1704         & 1746         \\
8        & 1531           & 1580           & 1719           & 1073         & 1105         & 1215         \\
10       & 1250           & 1344           & 1479           & 909          & 980          & 1022         \\
16       & 821            & 914            & 1031           & 630          & 714          & 736          \\
20       & 701            & 823            & 903            & 531          & 663          & 701          \\ \hline
\end{tabular}
\caption{Runtime (in seconds) of MDS on $100,000 \times 1000$ simulated data on multiple AWS EC2 instances} \label{tab:mdsaws}
\end{table}

\subsection[L1-regularized Cox regression]{$\ell_1$-regularized Cox regression}\label{sec:cox}

Another example we consider is the Cox proportional hazards model-based regression \citep{cox1972regression} with $\ell_1$-regularization. In this regression problem, the covariate matrix $X \in \mathbb{R}^{m \times n}$ is given, as well as the survival time for each sample, which is possibly right-censored. They are denoted by $y = (y_1, \dotsc, y_m)$ where $y_i = \min\{t_i, c_i\}$, and $t_i$ is the time to event and $c_i$ is time to right censoring. Let $\delta_i = I_{\{t_i \le c_i\}}$ indicate whether the event occured before  the right censoring of sample $i$. The log-partial likelihood of the Cox proportional hazards model is defined by 
\begin{align*}
L(\beta) &= \sum_{i=1}^m \delta_i \left[ \beta^\top x_i \log \left( \sum_{j: y_j \ge y_i } \exp(\beta^\top x_j)\right)\right].
\end{align*}
We maximize $L(\beta) - \lambda \|\beta\|_1$. 
Applying the proximal gradient algorithm, which is equivalent to the iterative soft-thresholding algorithm \citep{Beck:ISTA} to this problem, the iteration is given by:
\begin{align*} 
w_{(n+1)i} &= \exp(x_i^\top \beta); \;\; W_{(n+1)j} = \sum_{i: y_i \ge y_j}w_{(n+1)i} \\
\pi_{(n+1)ij} &= I(t_i \ge t_j) w_{(n+1)i}/W_{(n+1)j} \\
\Delta_{(n+1)} &= X^\top (I - P_{(n+1)}) \delta,\text{ where $P_{(n+1)} = (\pi_{(n+1)ij})$} \\
\beta_{(n+1)} &= \mathcal{S}_\lambda (\beta_{(n)} + \sigma \Delta_{(n+1)}),
\end{align*}
where $\mathcal{S}_\lambda(\cdot)$ is the soft-thresholding operator discussed in Section \ref{sec:dot}. If the samples are sorted in nonincreasing order of $y_i$, the summation $\sum_{i: y_i \ge y_j}w_{(n+1)i}$ can be implemented using the function \code{cumsum()}.
Convergence of the algorithm is guaranteed if we choose the step size $\sigma=1/(2\|X\|_2^2)$ \citep{ko2020high}.
A simple implementation of this algorithm in \proglang{Julia}, assuming no ties in $y_i$, can be written as:
\begin{verbatim}
using DistStat, LinearAlgebra, Random
m = 1000; n = 1000
T = Float64; A = Array; iter=1000
sigma = 0.00001 # step size
lambda = 0.00001 # penalty param
X = MPIMatrix{T,A}(undef, m, n); randn!(X)
delta = convert(A{T}, rand(m) .> 0.3) # 30% 1, 70% 0.
DistStat.Bcast!(delta) # bcast the delta from rank-0 proc.
y = convert(A{T}, collect(reverse(1:size(X,1))))
# assume decreasing observed time
beta = MPIVector{T,A}(undef, n)
pi_ind = MPIMatrix{T,A}(undef, m, m)
y_dist = distribute(reshape(y, 1, :))
fill!(pi_ind, one(T))
pi_ind .= ((pi_ind .* y_dist) .- y) .<= 0
Xbeta = A{T}(undef, m)
W = A{T}(undef, m)
pi_delta = A{T}(undef, m)
gradient = MPIVector{T, A}(undef, n)
for i in 1:iter
    mul!(Xbeta, X, beta)
    w = exp.(Xbeta)
    cumsum!(W, w)
    W_dist = distribute(reshape(W, 1, :))
    pi = pi_ind .* w ./ W_dist
    mul!(pi_delta, pi, delta)
    dmpd = delta .- pi_delta
    mul!(gradient, transpose(X), dmpd)
    beta .= soft_threshold.(beta .+ sigma .* gradient, lambda)
end
\end{verbatim}

For performance optimization, note that in addition to the memory for $X$, an intermediate storage for two $m \times m$ matrices are needed for \code{pi} and \code{pi\_ind}. This can be avoided by environment-specific implementation. For example, 
a CPU function to compute $P_{(n+1)} \delta$ can be written using \pkg{LoopVectorization.jl} \citep{loopvectorizationjl} for efficient SIMD parallelization using the Advanced Vector Extensions \citep[AVX,][]{firasta2008intel}. 
For GPU, a kernel directly computing $P_{(n+1)} \delta$ can be written using \pkg{CUDA.jl}. These environment-specific implementations not only use less memory, but also results in some speed-up. On the local workstation we used, the device-specific CPU implementation with four processes with each process using a single core took roughly half the time compared to the dot broadcasting-based implementation. The GPU implementation with four GPUs was 5-10\% faster. Code for accelerating computation of $P_{(n+1)} \delta$ is avialable in Appendix \ref{sec:coxcode}.

Table \ref{tab:coxgpu} demonstrates the scalability of the proximal gradient algorithm for $\ell_1$-regularized Cox regression on multiple GPUs. The speed-up with 8 GPUs is clear. While \pkg{DistStat.jl} was slightly slower than \pkg{dist\_stat} with fewere GPUs, the gap is in general small. 
Unfortunately, the underlying algorithm for the \code{cumsum()} method in \pkg{PyTorch} is numerically unstable, and could not be used for very small values of $\lambda$. For this reason, \citet{ko2020high} resorted to double precision. On the other hand, the \code{cumsum()} function from \pkg{CUDA.jl} is numerically stable for these values of $\lambda$.

\begin{table}[]
\centering
\begin{tabular}{rrrr} \hline
\multirow{2}{*}{GPUs}            & \pkg{dist\_stat}      & \multicolumn{2}{c}{\pkg{DistStat.jl}}      \\
        \cmidrule(lr){2-2} \cmidrule(lr){3-4}
 & \code{Float64} & \code{Float64} & \code{Float32} \\ \hline
1       & 382                        & 447                      & 292                      \\
2       & 205                        & 196                      & 113                      \\
3       & --                         & 160                      & 91                       \\
4       & 115                        & 136                      & 80                       \\
5       & 98                         & 121                      & 75                       \\
6       & --                         & 113                      & 71                       \\
7       & --                         & 106                      & 69                       \\
8       & 124                        & 86                       & 67                       \\ \hline
\end{tabular}
\caption{Runtime (in seconds) of $\ell_1$-regularized Cox regression on $10,000 \times 10,000$ simulated data on multiple GPUs with $\lambda = 10^{-8}$. Experiments  could not be run with 3, 6, and 7 GPUs on \pkg{dist\_stat}, because 3, 6, and 7 did not divide the data size.} \label{tab:coxgpu}
\end{table}

For the AWS experiments on \pkg{DistStat.jl}, 36 processes per instance was used once again. Table \ref{tab:coxaws} shows the runtime of the algorithm for 1000 iterations with a simulated $100,000 \times 200,000$ dataset. Thanks to the flexibility of the \proglang{Julia} implementation allowing 36 threads per instance, the speed-up of \pkg{DistStat.jl} over \pkg{dist\_stat}, which used two threads per instance, is obvious.

\begin{table}[]
\centering
\begin{tabular}{rrr} \hline
Nodes & \pkg{dist\_stat} & \pkg{DistStat.jl} \\ \hline
4        & 1455    & 918   \\
5        & 1169    & 819   \\
8        & 809     & 558   \\
10       & 618     & 447   \\
16       & 389     & 290   \\
20       & 318     & 245   \\ \hline
\end{tabular}
\caption{Runtime (in seconds) of $\ell_1$-regularized Cox regression on $100,000 \times 200,000$ simulated data on multiple AWS EC2 instances with $\lambda = 10^{-8}$.} \label{tab:coxaws}
\end{table}

\section{Genome-wide survival analysis of the UK Biobank dataset}\label{sec:ukbiobank}
In this section, we demonstrate an application of \pkg{DistStat.jl} on the genome-wide survival analysis on the UK Biobank dataset \citep{ukbiobank2015} using the $\ell_1$-regularized Cox regression code for Type 2 Diabetes (T2D).  
The UK Biobank dataset consists of 500,000 subjects and approximately 800,000 single nucleotide polymorphisms (SNPs). We used 402,297 patients including 17,994 T2D patients and 470,194 SNPs, after filtering possible Type 1 Diabetes patients and filtering SNPs with high genotyping failure rates and low minor allele frequencies. In addition to the SNPs, gender, and the top ten principal components were included as unpenalized covariates. Unlike \citet{ko2020high}, who used only a half of the full dataset due to the limitations of \pkg{dist\_stat}, the improved memory efficiency of \pkg{DistStat.jl} allowed us to use the entire dataset, which is about 350 gigabytes, with the additional memory for computation.
The age of onset was used as the ``survival time'' $y_i$ for T2D patients ($\delta = 1$), and the age at the last visit was used as $y_i$ for non-T2D subjects ($\delta = 0$). We used 43 different values of $\lambda$ in range $[6.0 \times 10^{-9}, 1.5 \times 10^{-8}]$, where 0 to 320 SNPs were selected. Breslow's method \citep{breslow1972discussion} was applied for any ties in $y_i$. We used 20 \code{c5.18xlarge} instances for the analysis. It took less than 2050 iteration until convergence, where convergence was determined by testing if $\frac{|f(\beta_{(n)}) - f(\beta_{(n - 10)})|}{|f(\beta_{(n)})+1|} < 10^{-5}$. For each $\lambda$, the experiment took between 3180 and 3720 seconds. 

We ranked the SNPs based on the largest of $\lambda$ for which each SNP has nonzero coefficient with ties broken by the absolute values of the coefficients when joining the model. The set of top nine selections was identical to that of \citet{ko2020high} with a slightly different order, as listed in Table \ref{tab:ukbk_topnine}. The set includes SNPs on TCF7L2, which is a well-established T2D-related gene \citep{scott2007genome,wellcome2007genome}, and those from SLC45A2 and HERC2, whose variants are known to affect skin, eye, and hair pigmentation \citep{cook2009analysis}. Occurrence of SLC45A2 and HERC2 may be because of bias in European subjects. As with \citet{ko2020high}, significance test using unpenalized Cox regression with only selected SNPs, gender, and top 10 principal components gave a result that is more specific to T2D.
We selected SNPs with $p$-values less than $0.01/333$ using Bonferroni correction to control the family-wise error rate under 0.01. Table \ref{tab:ukbk_significant} lists the 9 selected SNPs. 

Six of the SNPs, including the SNPs with five lowest $p$-values were previously reported to have direct association with T2D (rs1801212 from WFS1 \citep{fawcett2010detailed}, rs4506565 from TCF7L2 \citep{wellcome2007genome,dupuis2010new}, rs2943640 from IRS1 \citep{langenberg2014gene}, rs10830962 from MTNR1B \citep{klimentidis2014identification,salman2015mtnr1b}, rs343092 from HMGA2 \citep{ng2014meta}, and rs231362 from KCNQ1 \citep{riobello2016kcnq1}). In addition, rs1351394 is from HMGA2, known to be associated with T2D. This appears to be an improvement over \citet{ko2020high} in which three of the top nine selections were found to be directly associated with T2D and three others were on the known T2D-associated genes.

\begin{table}[ht!]
\centering
\begin{threeparttable}
\resizebox{\textwidth}{!}{
\begin{tabular}{rlrrrrrlr} \hline
Rank & SNP ID     & Chr & Location   & A1\tnote{A} & A2\tnote{B} & MAF\tnote{C}          & Mapped Gene & Sign \\ \hline
1    & rs4506565  & 10  & 114756041  & A  & \textbf{T}  & 0.314 & TCF7L2      & $+$    \\
2    & rs16891982 & 5   & 33951693   & G  & \textbf{C}  & 0.073 & SLC45A2     & $-$    \\
3    & rs12243326 & 10  & 114788815  & T  & \textbf{C}  & 0.281 & TCF7L2      & $+$    \\
4    & rs12255372 & 10  & 1148088902 & G  & \textbf{T}  & 0.285 & TCF7L2      & $+$    \\
5    & rs28777    & 5   & 33958959   & A  & \textbf{C}  & 0.062 & SLC45A2     & $-$    \\
6    & rs35397    & 5   & 33951116   & T  & \textbf{G}  & 0.096 & SLC45A2     & $-$    \\
7    & rs1129038  & 15  & 28356859   & T  & \textbf{C}  & 0.261 & HERC2       & $-$    \\
8    & rs12913832 & 15  & 28365618   & G  & \textbf{A}  & 0.259 & HERC2       & $-$    \\
9    & rs10787472 & 10  & 114781297  & A  & \textbf{C}  & 0.470 & TCF7L2      & $+$    \\ \hline
\end{tabular}
}
\begin{tablenotes}
\small \item[A] Major allele, \item[B] Minor allele, \item[C] Minor allele frequency. The boldface indicates the risk allele determined by the reference allele and the sign of the regression coefficient. 
\end{tablenotes}
\end{threeparttable}
\caption{Top nine SNPs selected by $\ell_1$-penalized Cox regression}\label{tab:ukbk_topnine}
\end{table}

\begin{table}[ht!]
\centering
\resizebox{\textwidth}{!}{
\begin{threeparttable}
\begin{tabular}{lrrrrrlrr} \hline
SNP ID     & Chr & Location  & A1\tnote{A} & A2\tnote{B} & MAF\tnote{C}   & Mapped Gene & Coefficient & $p$-value        \\ \hline
rs1801212  & 4   & 6302519   & \textbf{A}  & G  & 0.270 & WFS1        & 0.1123       & \textless{}2E-16 \\
rs4506565  & 10  & 114756041 & A  & \textbf{T}  & 0.314 & TCF7L2      & 0.2665       & \textless{}2E-16 \\
rs2943640  & 2   & 227093585 & \textbf{C}  & A  & 0.336 & IRS1           & 0.0891       & 1.57E-14         \\
rs10830962 & 11  & 92698427  & C  & \textbf{G}  & 0.402 & MTNR1B      & 0.0731       & 1.46E-11         \\
rs343092   & 12  & 66250940  & G  & \textbf{T}  & 0.166 & HMGA2       & -0.0746      & 2.26E-07         \\
rs1351394  & 12  & 66351826  & \textbf{C}  & T  & 0.478 & HMGA2       & 0.0518       & 1.70E-06         \\ 
rs2540917  & 2   & 60608759  & \textbf{T}  & C  & 0.389 & RNU1-32P    & -0.0476      & 2.18E-05         \\
rs1254207  & 1   & 236368227 & C  & \textbf{T}  & 0.395 & GPR137B     & 0.0458       & 2.84E-05         \\
rs231362   & 11  & 2691471   & \textbf{G}  & A  & 0.461 & KCNQ1       & 0.0607       & 2.87E-05         \\
\hline
\end{tabular}
\begin{tablenotes}
\small \item[A] Major allele, \item[B] Minor allele, \item[C] Minor allele frequency. The boldface indicates the risk allele determined by the reference allele and the sign of the regression coefficient. 
\end{tablenotes}
\end{threeparttable}
}
\caption{SNPs with significant coefficients with significance level 0.01 after Bonferroni correction}\label{tab:ukbk_significant}
\end{table}

\section{Summary and Discussion}\label{sec:summary}

\pkg{DistStat.jl} is an initial step toward a unified programming in various HPC environments, which supplies a distributed array data structure with any underlying array, essential for high-performance distributed statistical computing. While tested on the basic \code{Array} on CPU nodes and \code{CuArray} on multi-GPU workstations, \pkg{DistStat.jl} can be used with any array type on any HPC environment provided that the array interface is implemented in \proglang{Julia} with \pkg{MPI} support. 

Statistical applications of \pkg{DistStat.jl} including NMF, MDS, and $\ell_1$-regularized Cox regularization is examined, and scalability are demonstrated on a 8-GPU workstation and a virtual cluster on AWS cloud with up to 20 nodes. The performance was equivalent to or better than its \proglang{Python} counterpart, \pkg{dist\_stat}: the computation was 20-30\% faster on CPUs because of flexibility in setting number of threads per node, and with unoptimized GPU kernels, performance on the multi-GPU workstation was comparable to \pkg{dist\_stat}. Using \pkg{DistStat.jl}, we were able to analyze a large genomic dataset of size $400,000 \times 500,000$, which appears the largest analyzed using a multivariate survival model. (Conventional genome-wide analysis relies on SNP-by-SNP univariate models.)  

\pkg{DistStat.jl} can be extended to communication-avoiding linear algebra \citep{ballard2011minimizing}. For example, an interface to \pkg{ScaLAPACK} \citep{choi1992scalapack} can be merged into \pkg{DistStat.jl} for communication-avoiding dense matrix multiplication for CPU. For sparse matrix-dense matrix multiplication \citep{koanantakool2016communication} often used in inverse covariance matrix estimation for graphical models \citep{koanantakool2018communication}, its incorporation into \pkg{DistStat.jl} will provide a highly productive distributed CPU and multi-GPU implementations at the same time. Furthermore, the distributed array can be used as the backbone for implementation of communication-avoiding statistical inference \citep{jordan2019communication} or applications such as communication-avoiding least angle regression \citep{fountoulakis2019parallel}.

\appendix

\section{Installation} \label{sec:install}
\pkg{DistStat.jl} requires a standard installation of \pkg{MPI} as a prerequisite.  
With \pkg{MPI} prepared, the rest of installation follows the standard method for GitHub package in \proglang{Julia}:
\begin{verbatim}
julia> using Pkg
julia> pkg"add https://github.com/kose-y/DistStat.jl"
\end{verbatim}
Then, all the dependencies are installed along with the package \pkg{DistStat.jl}.

For the optional multi-GPU support, \pkg{MPI} should be installed with ``\pkg{CUDA}-aware'' support. For example, when using \pkg{OpenMPI} \citep{graham2006open}, it should be configured with the flag \code{----with-cuda}. In the shell: 
\begin{verbatim}
$ ./configure --with-cuda
$ make install
\end{verbatim}
From the \proglang{Julia} side, the standard set of \pkg{CUDA} interface package, \pkg{CUDA.jl} is required. 
\pkg{MPI.jl}, the underlying \pkg{MPI} interface, supports \pkg{CUDA}-aware \pkg{MPI}. Set the environment variable \code{JULIA\_MPI\_BINARY=system} with \pkg{CUDA}-aware \pkg{MPI} as the system-wide \pkg{MPI}.

\section[Code for memory-efficient l1-regularized Cox proportional hazards model]{Code for memory-efficient $\ell_1$-regularized Cox proportional hazards model} \label{sec:coxcode}
For CPU code, the following code accelerates the computation of $P_{(n+1)} \delta$ for $\ell_1$-regularized Cox regression in Section \ref{sec:cox} using the AVX.
\begin{verbatim}
using LoopVectorization
function pi_delta!(out, w, W_dist, delta, W_range)
    # fill `out` with zeros beforehand.
    m = length(delta)
    W_base = minimum(W_range) - 1
    W_local = W_dist.localarray
    @avx for i in 1:m
        outi = zero(eltype(w))
        for j in 1:length(W_range)
            outi += ifelse(i <= j + W_base, 
                delta[j + W_base] * w[i] / W_local[j], zero(eltype(w)))
        end
        out[i] = outi
    end
    DistStat.Barrier()
    DistStat.Allreduce!(out)
    return out
end
\end{verbatim}
\code{DistStat.Allreduce!(out)} computes the elementwise sum of \code{out} in all ranks, and saves it in the place of \code{out}. For GPU, the kernel function can be written as follows:
\begin{verbatim}
function pi_delta_kernel!(out, w, W_dist, delta, W_range)
    idx_x = (blockIdx().x-1) * blockDim().x + threadIdx().x
    stride_x = blockDim().x * gridDim().x
    W_base = minimum(W_range) - 1
    for i in idx_x:stride_x:length(out)
        for j in W_range
            @inbounds if i <= j
                out[i] += delta[j] * w[i] / W_dist[j - W_base]
            end
        end
    end  
end
\end{verbatim}
And the host function to compute $P_{(n+1)} \delta$ is:
\begin{verbatim}
function pi_delta!(out::CuArray, w::CuArray, W_dist, delta, W_range)
    fill!(out, zero(eltype(out)))
    numblocks = ceil(Int, length(w)/256)
    CUDA.@sync begin
        @cuda threads=256 blocks=numblocks pi_delta_kernel!(out, 
            w, W_dist.localarray, delta, W_range)
    end
    DistStat.Allreduce!(out)
    out
end
\end{verbatim}
\bibliography{refs}
\bibliographystyle{apalike}
\end{document}